\documentclass{llncs}

\usepackage{color}
\usepackage{comment}
\usepackage{amsmath}
\usepackage{amssymb}
\usepackage{enumitem}
\usepackage[ruled,lined,linesnumbered,boxed,vlined]{algorithm2e}
\usepackage{tikz}
\usepackage{pgfplots}
\usetikzlibrary{decorations.pathmorphing}
\usetikzlibrary{decorations.pathreplacing}
\usetikzlibrary{patterns}
\usetikzlibrary{arrows.meta}
\usetikzlibrary{shapes.misc}
\usetikzlibrary{matrix}
\usepgfplotslibrary{groupplots}

\usepackage{subfig}

\newcommand{\BF}[0]{\ensuremath{\texttt{SSGS}^\texttt{BF}}}
\newcommand{\NBF}[0]{\ensuremath{\texttt{SSGS}^\texttt{NBF}}}
\newcommand{\data}[0]{\ensuremath{\texttt{SSGS}^\texttt{data}}}
\newcommand{\Conv}[0]{\ensuremath{\texttt{SSGS}^\texttt{conv}}}
\newcommand{\Hybrid}[0]{\ensuremath{\texttt{Hybrid}}}

\SetKwInOut{Input}{input}
\SetKwInOut{Output}{output}
\SetKw{GotoLine}{go to line}

\begin{document}

\title{Efficient Adaptive Implementation of the Serial Schedule Generation Scheme using Preprocessing and Bloom Filters}
\titlerunning{Hamiltonian Mechanics}  
%
\author{Daniel Karapetyan\inst{1} \and Alexei Vernitski\inst{2}}
\authorrunning{D. Karapetyan and A. Vernitsky.} 
\institute{Institute for Analytics and Data Science, University of Essex, Essex, UK\\
\email{daniel.karapetyan@gmail.com}
\and
Department of Mathematical Sciences, University of Essex, Essex, UK\\
\email{asvern@essex.ac.uk}}

\maketitle              

\begin{abstract}
 The majority of scheduling metaheuristics use indirect representation of solutions as a way to efficiently explore the search space.
 Thus, a crucial part of such metaheuristics is a ``schedule generation scheme'' -- procedure translating the indirect solution representation into a schedule.
 Schedule generation scheme is used every time a new candidate solution needs to be evaluated.
 Being relatively slow, it eats up most of the running time of the metaheuristic and, thus, its speed plays significant role in performance of the metaheuristic.
 Despite its importance, little attention has been paid in the literature to efficient implementation of schedule generation schemes.
 We give detailed description of serial schedule generation scheme, including new improvements, and propose a new approach for speeding it up, by using Bloom filters.
 The results are further strengthened by automated control of parameters.
 Finally, we employ online algorithm selection to dynamically choose which of the two implementations to use.
 This hybrid approach significantly outperforms conventional implementation on a wide range of instances.


\end{abstract}
\section{Introduction}

 Resource Constrained Project Scheduling Problem (RCPSP) is to schedule a set of jobs $J$ subject to precedence relationships and resource constraints.
 RCPSP is a powerful model generalising several classic scheduling problems such as job shop scheduling, flow shop scheduling and parallel machine scheduling.
 
 In RCPSP, we are given a set of resources $R$ and their capacities $c_r$, $r \in R$.
 In each time slot, $c_r$ units of resource $r$ are available and can be shared between jobs.
 Each job $j \in J$ has a prescribed consumption $v_{j,r}$ of each resource $r \in R$.
 We are also given the duration $d_j$ of a job $j \in J$.
 A job consumes $v_{j,r}$ units of resource $r$ in every time slot that it occupies.
 Once started, a job cannot be interrupted (no preemption is allowed).
 Finally, each resource is assigned a set $\mathit{pred}_j \subset J$ of jobs that need to be completed before $j$ can start.
 
 There exist multiple extensions of RCPSP\@.
 In the multi-mode extension, each job can be executed in one of several modes, and then resource consumption and duration depend on the selected mode.
 In some applications, resource availability may vary with time.
 There could be set-up times associated with certain jobs.
 In multi-project extension, several projects run in parallel sharing some but not all resources.
 In this paper we focus on the basic version of RCPSP, however some of our results can be easily generalised to many of its extensions and variations.

 Most of the real-world scheduling problems, including RCPSP, are NP-hard, and hence only problems of small size can be solved to optimality, whereas for larger problems (meta)heuristics are commonly used.
 Metaheuristics usually search in the space of feasible solutions; with a highly constrained problem such as RCPSP, browsing the space of feasible solutions is hard.
 Indeed, if a schedule is represented as a vector of job start times, then changing the start time of a single job is likely to cause constraint violations.
 Usual approach is to use indirect solution representation that could be conveniently handled by the metaheuristic but could also be efficiently translated into the direct representation.
 
 Two translation procedures widely used in scheduling are \emph{serial schedule generation scheme} (SSGS) and \emph{parallel schedule generation scheme}~\cite{Kolisch1999}.
 Some studies conclude that SSGS gives better performance~\cite{Kim2010}, while others suggest to employ both procedures within a metaheuristic~\cite{Kolisch2006}.
 Our research focuses on SSGS.

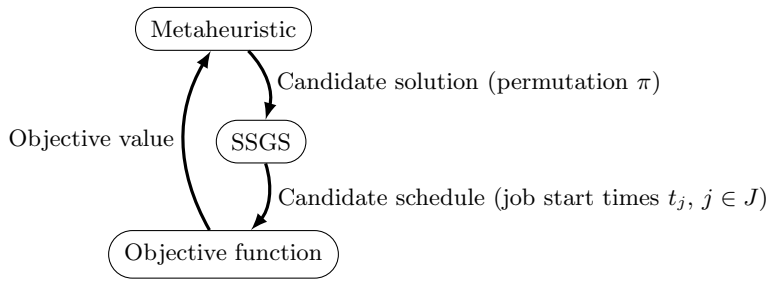
\begin{figure}[htb]
	\centering
	\begin{tikzpicture}[
	    component/.style={rounded rectangle, draw, inner sep=5pt},
        arrow/.style={-latex, very thick}
   	]
	    \node[component] (MH) at (0, 3) {Metaheuristic};
	    \node[component] (SSGS) at (0.5, 1.5) {SSGS};
	    \node[component] (OF) at (0, 0) {Objective function};
        
        \path[arrow]
        	(MH) edge[bend left] node[anchor=west] {Candidate solution (permutation $\pi$)} (SSGS);
        \path[arrow]
            (SSGS) edge[bend left] node[anchor=west] {Candidate schedule (job start times $t_j$, $j \in J$)} (OF);

        \path[arrow]
            (OF) edge[bend left] node[anchor=east] {Objective value} (MH);
	\end{tikzpicture}

	\caption{
    	Classic architecture of a scheduling metaheuristic.
        To obtain objective value of a candidate solution, metaheuristic uses SSGS to translate the candidate solution into a candidate schedule, which is then used by objective function.
    }
    \label{fig:architecture}
\end{figure}

 With SSGS, the indirect solution representation is a permutation $\pi$ of jobs.
 The metaheuristic handles candidate solutions in indirect (permutation) form.
 Every time a candidate solution needs to be evaluated, SSGS is executed to translate the solution into a schedule ($t_j$, $j \in J$), and only then the objective value can be computed, see Figure~\ref{fig:architecture}.

 SSGS is a simple procedure that iterates through $J$ in the order given by $\pi$, and schedules one job at a time, choosing the earliest feasible slot for each job.
 The only requirement for $\pi$ is to respect the precedence relations; otherwise SSGS produces a feasible schedule for any permutation of jobs.
 The pseudo-code of SSGS is given in Algorithm~\ref{alg:ssgs}, and its two subroutines $\mathit{find}$ and $\mathit{update}$ in Algorithms~\ref{alg:naive} and~\ref{alg:update}.

\begin{algorithm}[tb]
	\caption{Serial Schedule Generation Scheme (SSGS)\@.  Here $T$ is the upper bound on the makespan.
    \label{alg:ssgs}}
    
	\Input {Solution $\pi$ in permutation form, respecting precedence relations}
	\Output {Schedule $t_j$, $j \in J$}
    
    $A_{t, r} \gets c_r$ for every $t = 1, 2, \ldots, T$ and $r \in R$\;  \label{line:initialise-a}
    \For{$i = 1, 2, \ldots, |J|$}
    {
    	$j \gets \pi(i)$\;
        $t^0 \gets \max_{j' \in \mathit{pred}_j} t_{j'} + d_{j'}$\;
    	$t_j \gets \mathit{find}(j, t^0, A)$ (see Algorithm~\ref{alg:naive})\; 
        $\mathit{update}(j, t_j, A)$ (see Algorithm~\ref{alg:update})\;
    }

	\Return{$t_j$, $j \in J$}\;
\end{algorithm} 

\begin{algorithm}[tb]
	\caption{Conventional implementation of $\mathit{find}(j, t^0, A)$ -- a function to find the earliest feasible slot for job $j$.
    \label{alg:naive}}
    
	\Input {Job $j \in J$ to be scheduled}
	\Input {Earliest start time $t^0$ as per precedence relations}
	\Input {Current availability $A$ of resources}
	\Output {Earliest feasible start time for job $j$}
     
    
    $t_j \gets t^0$\;
    $t \gets t_j$\;
    \While {$t < t_j + d_j$}
    {
       	\uIf {$A_{t, r} \ge v_{j, r}$ for every $r \in R$}
        {
          	$t \gets t + 1$\;
        }
    	\Else
        {
        	$t_j \gets t + 1$\;
            $t \gets t_j$\;
		}
    }
    

	\Return{$t_j$}\;
\end{algorithm} 

\begin{algorithm}[tb]
	\caption{Procedure $\mathit{update}(j, t_j, A)$ to update resource availability $A$ after scheduling job $j$ at time $t_j$.
    \label{alg:update}}
    
	\Input {Job $j$ and its start time $t_j$}
	\Input {Resource availability $A$}

	\For {$t \gets t_j, t_j + 1, \ldots, t_j + d_j - 1$\label{line:update-loop-slots}}
    {
		$A_{j, r} \gets A_{j, r} - v_{j, r}$ for every $r \in R$\;  \label{line:update-loop-resources}
	}
\end{algorithm}


 Commonly, the objective of RCPSP is to find a schedule that minimises the makespan, i.e.\ the time required to complete all jobs; however other objective functions are also considered in the literature.
 We say that an objective function of a scheduling problem is \emph{regular} if advancing the start time of a job cannot worsen the solution's objective value.
 Typical objective functions of RCPSP, including makespan, are regular.
 If the scheduling problem has a regular objective function, then SSGS guarantees to produce \emph{active} solutions, i.e.\ solutions that cannot be improved by changing $t_j$ for a single $j \in J$.
 Moreover, it was shown~\cite{Kolisch1996} that for any active schedule $S$ there exists a permutation $\pi$ for which SSGS will generate $S$.
 Since any optimal solution $S$ is active, searching in the space of feasible permutations $\pi$ is sufficient to solve the problem.
 This is an important property of SSGS; the parallel schedule generation scheme, mentioned above, does not provide this guarantee \cite{Kolisch1996} and, hence, may not in some circumstances allow a metaheuristic finding optimal or near-optimal solutions.


 The runtime of a metaheuristic is divided between its search control mechanism that modifies solutions and makes decisions such as accepting or rejecting new solutions, and SSGS\@.
 While SSGS is a polynomial algorithm, in practice it eats up the majority of the metaheuristic runtime (over 98\% as reported in~\cite{MISTA}).
 In other words, by improving the speed of SSGS twofold, one will (almost) double the number of iterations a metaheuristic performs within the same time budget, and this increase in the number of iterations is likely to have a major effect on the quality of obtained solutions.
 
 In our opinion, not enough attention was paid to SSGS -- a crucial component of many scheduling algorithms, and this study is to close this gap.
 In this paper we discuss approaches to speed up the conventional implementation of SSGS\@.
 Main contributions of our paper are:
\begin{itemize}
	\item
    A detailed description of SSGS including old and new speed-ups (Section~\ref{sec:ssgs-nbf}).

    \item
    New implementation of SSGS employing Bloom filters for quick testing of resource availability (Section~\ref{sec:bloom-filters}).

    \item
    A hybrid control mechanism that employs intelligent learning to dynamically select the best performing SSGS implementation (Section~\ref{sec:hybrid}).
\end{itemize}
 
 Empirical evaluation in Section~\ref{sec:evaluation} confirms that both of our implementations of SSGS perform significantly better than the conventional SSGS, and the hybrid control mechanism is capable of correctly choosing the best implementation while generating only negligible overheads.

 

\section{SSGS implementation details}
\label{sec:ssgs-nbf}

 Before we proceed to introducing our main new contributions in Sections~\ref{sec:bloom-filters} and~\ref{sec:hybrid}, we describe what state-of-the-art implementation of SSGS we use, including some previously unpublished improvements.
 
\subsection{Initialisation of $A$}
 
 The initialisation of $A$ in line~\ref{line:initialise-a} of Algorithm~\ref{alg:ssgs} iterates through $T$ slots, where $T$ is the upper bound on the makespan.
 It was noted in~\cite{MISTA} that instead of initialising $A$ at every execution of SSGS, one can reuse this data structure between the executions. 
 To correctly initialise $A$, at the end of SSGS we restore $A_{t, r}$ for each $r \in R$ and each slot where some job was scheduled: $A_{t, r} \gets c_r$ for $r \in R$ and $t = 1, 2, \ldots, M$, where $M$ is the makespan of the solution.
 Since $M \le T$ and usually $M \ll T$, this notably improves the performance of SSGS~\cite{MISTA}.




\subsection{Efficient search of the earliest feasible slot for a job}
\label{sec:find-enhancements}

 The function $\mathit{find}(j, t^0, I, A)$  finds the earliest slot feasible for scheduling job $j$. 
 Its conventional implementation (Algorithm~\ref{alg:naive}) takes $O(T |R|)$ time, where $T$ is the upper bound of the time horizon.
 Our enhanced implementation of $\mathit{find}$ (Algorithm~\ref{alg:find}), first proposed in~\cite{MISTA}, has the same worst case complexity but is more efficient in practice.
 It is inspired by the Knuth-Morris-Pratt substring search algorithm.
 Let $t_j$ be the assumed starting time of job $J$.
 To verify if it is feasible, we need to test sufficiency of resources in slots $t_j, t_j + 1, \ldots, t_j + d_j - 1$.
 Unlike the conventional implementation, our enhanced version tests these slots in the reversed order.
 The order makes no difference if the slot is feasible, but otherwise testing in reversed order allows us to skip some slots; in particular, if slot $t$ is found to have insufficient resources then we know that feasible $t_j$ is at least $t + 1$.
 
 A further speed up, which was not discussed in the literature before, is to avoid re-testing of slots with sufficient resources.
 Consider the point when we find that the resources in slot $t$ are insufficient.
 By that time we know that the resources in $t + 1, t + 2, \ldots, t_j + d_j - 1$ are sufficient.
 Our heuristic is to remember that the earliest slot $t^\text{test}$ to be tested in future iterations is $t_j + d_j$.
 


\newcommand{\cellfont}[0]{\small\bfseries}

\tikzset{
	cell base/.style={minimum width=1cm, minimum height=1cm, draw=none, transform shape},
	unknown/.style={cell base, fill=white!90!black, label={[font=\cellfont]center:?}},
	available/.style={cell base, fill=none, label={[font=\cellfont]center:$+$}},
	unavailable/.style={cell base, fill=none, label={[font=\cellfont]center:$-$}},
    known available/.style={cell base, pattern=north east lines, pattern color=gray,label={[font=\cellfont]center:$+$}}
}

\newcommand{\slots}[4]{
	\xdef\x{0}
	\xdef\y{#2}

   	\node[anchor=west] at (-6,0.5+\y) {#3};

	\foreach \cell in #1
	{
		\node[\cell] at (\x + 0.5, 0.5+\y) {};
		\pgfmathparse{\x+1}
		\xdef\x{\pgfmathresult}
	}

	\xdef\x{0}
	\foreach \cell in #1
	{
		\draw[black] (\x,\y) -- (\x,1+\y);
		\pgfmathparse{\x+1}
		\xdef\x{\pgfmathresult}
	}

	\draw[black] (\x,\y) -- (\x,1+\y);
	\draw[black,decorate, decoration=snake] (0.5+\x,\y) -- (0.5+\x,1+\y);
    \draw[black] (0,\y) -- (0.5+\x,\y);
    \draw[black] (0,1+\y) -- (0.5+\x,1+\y);
    

	\ifthenelse{#4 > -1}{\draw[ultra thick] (#4,\y) rectangle (4+#4,1+\y);}{}

}

\begin{algorithm}[tb]
	\caption{Enhanced implementation of $\mathit{find}(j, t^0, A)$
    \label{alg:find}}
    
	\Input {Job $j \in J$ to be scheduled}
	\Input {Earliest start time $t^0$ as per precedence relations}
	\Input {Current availability $A$ of resources}
	\Output {Earliest feasible start time for job $j$}
    
    $t_j \gets t^0$\;
	$t \gets t_j + d_j - 1$\;
	$t^\text{test} \gets t_j$\;
    \While {$t \ge t^\text{test}$}
    {
        \uIf {$A_{t, r} \ge V_{j, r}$ for every $r \in R$ \label{line:find-resources-loop}}
        {
            $t \gets t - 1$\;
        }
        \Else
        {
            $t^\text{test} \gets t_j + d_j$\;
            $t_j \gets t + 1$\;
            $t \gets t_j + d_j - 1$\;
        }
    }
        
	\Return{$t_j$}\;
\end{algorithm}

\subsection{Preprocessing and Automated Parameter Control}

 
 We observe that in many applications, jobs are likely to require only a subset of resources.
 For example, in construction works, to dig a hole one does not need cranes or electricians, hence the `dig a hole' job will not consume those resources.
 To exploit this observation, we pre-compute vector $R_j$ of resources used by job $j$, and then iterate only through resources in $R_j$ when testing resource sufficiency in $\mathit{find}$ (Algorithm~\ref{alg:find}, line~\ref{line:find-resources-loop}) and updating resource availability in $\mathit{update}$ (Algorithm~\ref{alg:update}, line~\ref{line:update-loop-resources}).
 Despite the simplicity of this idea, we are not aware of anyone using or mentioning it before.
 
 We further observe that some jobs may consume only one resource.
 By creating specialised implementations of $\mathit{find}$ and $\mathit{update}$, we can reduce the depth of nested loops.
 While this makes no difference from the theoretical point of view, in practice this leads to considerable improvement of performance.
 Correct implementations of $\mathit{find}$ and $\mathit{update}$ are identified during preprocessing and do not cause overheads during executions of SSGS\@.
 
 Having individual vectors $R_j$ of consumed resources for each job, we can also intelligently learn the order in which resource availability is tested (Algorithm~\ref{alg:find}, line~\ref{line:find-resources-loop}).
 By doing this, we are aiming at minimising the expected number of iterations within the resource availability test.
 For example, if resource $r$ is scarce and job $j$ requires significant amount of $r$, then we are likely to place $r$ at the beginning of $R_j$.
 More formally, we sort $R_j$ in descending order of probability that the resource is found to be insufficient during the search.
 This probability is obtained empirically by a special implementation of SSGS which we call \data{}.
 \data{} is used once to count how many times each resource turned out to be insufficient during scheduling of a job.
 (To avoid bias, \data{} tests every resource in $R_j$ even if the test could be terminated early.)
 After a single execution of \data{}, vectors $R_j$ are optimised, and in further executions default implementation of SSGS is used.
 
 
 One may notice that the data collected in the first execution of SSGS may get outdated after some time; this problem in addressed in Section~\ref{sec:hybrid}.

 Ordering of $R_j$ is likely to be particularly effective on instances with asymmetric use of resources, i.e.\ on real instances.
 Nevertheless, we observed improvement of runtime even on pseudo-random instances as reported in Section~\ref{sec:evaluation}.

\section{SSGS Implementation using Bloom filters}
\label{sec:bloom-filters}


 Performance bottleneck of an algorithm is usually its innermost loop.
 Observe that the innermost loop of the $\mathit{find}$ function is the test of resource sufficiency in a slot, see Algorithm~\ref{alg:find}, line~\ref{line:find-resources-loop}.
 In this section we try to reduce average runtime of this test from $O(|R|)$ to $O(1)$ time.
 For this, we propose a novel way of using a data structure known as Bloom filter.

 Bloom filters were introduced in~\cite{bloom1970space} as a way of optimising dictionary lookups, and found many applications in computer science and electronic system engineering~\cite{broder2004network,tarkoma2012theory}. 
 Bloom filters usually utilise pseudo-random hash functions to encode data, but in some applications~\cite{kayaturan2016way} non-hash-based approaches are used.
 In our paper, we also use a non-hash-based approach, and to our knowledge, our paper is the first in which the structure of Bloom filters is chosen dynamically according to the statistical properties of the data, with the purpose of improving the speed of an optimisation algorithm.
 
 In general, Bloom filters can be defined as a way of using data, and they are characterised by two aspects: first, all data is represented by short binary arrays of a fixed length (perhaps with a loss of accuracy); second, the process of querying data involves only bitwise comparison of binary arrays (which makes querying data very fast).

 We represent both each job's resource consumption and resource availability at each time slot, by binary arrays of a fixed length; we call these binary arrays Bloom filters.
 Our Bloom filters will consist of bits which we call \emph{resource level bits}.
 Each resource bit, denoted by $u_{r,k}$, $r \in R$, $k \in \{ 1, 2, \ldots, c_r \}$, means ``$k$ units of resource $r$'' (see details below). Let $U$ be the set of all possible resource bits.
 A \emph{Bloom filter structure} is  an ordered subset $L \subseteq U$, see Figure~\ref{fig:lossy-bf} for an example.
 Suppose that a certain Bloom filter structure $L$ is fixed.
 Then we can introduce $B^L(j)$, the Bloom filter of job $j$, and $B^L(t)$, the Bloom filter of time slot $t$, for each $j$ and $t$.
 Each $B^L(j)$ and $B^L(t)$ consists of $|L|$ bits defined as follows: if $u_{r,k}$ is the $i$th element of $L$ then
 $$
 B^L(j)_i = \begin{cases}
 	1 & \text{if } v_{j,r} \ge k, \\
    0 & \text{otherwise,}
 \end{cases}
 \qquad
 \text{and}
 \qquad
 B^L(t)_i = \begin{cases}
 	1 & \text{if } A_{t,r} \ge k, \\
    0 & \text{otherwise.}
 \end{cases}
 $$


\newcommand{\bfcellfont}[0]{\small\bfseries}

\tikzset{
	bf bit/.style={minimum width=1cm, minimum height=0.5cm, draw=black, transform shape, font=\small\bfseries},
    brace/.style={decorate,decoration={brace,amplitude=5pt,raise=1pt}}
}

\newcommand{\bloomfilter}[2]{
	\xdef\x{0}
	\xdef\y{#2}

	\foreach \bit in #1
	{
		\node[bf bit] at (\x + 0.5, \y - 0.25) {\bit};
		\pgfmathparse{\x+1}
		\xdef\x{\pgfmathresult}
	}
}

\newcommand{\belowbrace}[4]{\draw[brace] (#2,#3) -- (#1,#3) node[midway, below=1.5ex] {#4};}

\begin{figure}[tb]
\centering
	\begin{tikzpicture}[scale=1, y=-1cm]
		\xdef\x{1}
		\foreach \r/\i in {1/2, 1/3, 1/4, 2/1, 2/3, 3/1, 3/3, 3/4}
        {
			\node[anchor=south] at (\x - 0.5, -0.6) {$u_{\r,\i}$};
			\pgfmathparse{\x+1}
			\xdef\x{\pgfmathresult}
		}

		\bloomfilter{{$\ge 2$, $\ge 3$, $\ge 4$,   $\ge 1$, $\ge 3$,   $\ge 1$, $\ge 3$, $\ge 4$}}{0};

		\belowbrace{0}{3}{0}{Resource 1}
		\belowbrace{3}{5}{0}{Resource 2}
		\belowbrace{5}{8}{0}{Resource 3}
        
	\end{tikzpicture}

	\caption{
    	Example of a Bloom filter structure for a problem with 3 resources, each having capacity 4. 
    }
    \label{fig:lossy-bf}
\end{figure}
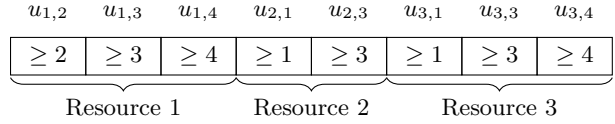 

 To query if a job $j$ can be scheduled in a time slot $t$, we compare Bloom filters $B^L(j)$ and $B^L(t)$ bitwise; then one of three situations is possible, as the following examples show.
 Consider a job $j$ and three slots, $t$, $t'$ and $t''$, with the following resource consumption/availabilities, and Bloom filter structure as in Figure~\ref{fig:lossy-bf}:
 \begin{alignat*}{7}
 v_{j,1} &= 3 \qquad & v_{j,2} &= 2 \qquad & v_{j,3} &= 0 \qquad & B^L(j) &= (110\,10\,000) \\[1ex]
 A_{t,1} &= 2 \qquad & A_{t,2} &= 3 \qquad & A_{t,3} &= 4 \qquad & B^L(t) &= (100\,11\,111) \\
 A_{t',1} &= 3 \qquad & A_{t',2} &= 1 \qquad & A_{t',3} &= 4 \qquad & B^L(t') &= (110\,10\,111) \\
 A_{t'',1} &= 3 \qquad & A_{t'',2} &= 2 \qquad & A_{t'',3} &= 2 \qquad & B^L(t'') &= (110\,10\,100)
 \end{alignat*}
 For two bit arrays of the same length, let notation `$\le$' mean bitwise less or equal.
By observing that $B^L(j) \not\le B^L(t)$, we conclude that resources in slot $t$ are insufficient for $j$; this conclusion is guaranteed to be correct.
 By observing that $B^L(j) \le B^L(t')$, we conclude tentatively that resources in slot $t'$ may be sufficient for $j$; however, further verification of the complete data related to $j$ and $t'$ (that is, $v_{j,\cdot}$ and $A_{t',\cdot}$) is required to get a precise answer; one can see that $v_{j,2} \ge A_{t',2}$, hence this is what is called a \emph{false positive}.
 Finally, we observe that $B^L(j) \le B^L(t'')$, and a further test (comparing $v_{j,\cdot}$ and $A_{t'',\cdot}$) confirms that resources in $t''$ are indeed sufficient for $j$.



 Values of $B^L(j)$, $j \in J$, are pre-computed when $L$ is constructed, and $B^L(t)$, $t = 1, 2, \ldots, T$, are maintained by the algorithm.

\subsection{Optimisation of Bloom Filter Structure}
 
 The length $|L|$ of a Bloom filter is limited to reduce space requirements and, more importantly for our application, speed up Bloom filter tests. 
 Note that if $|L|$ is small (such as 32 or 64 bits) then we can exploit efficient bitwise operators implemented by all modern CPUs; then each Bloom filter test takes only one CPU operation.
 We set $|L| = 32$ in our implementation.
 While obeying this constraint, we aim at minimising the number of false positives, because false positives slow down the implementation.
 
 
 
\begin{figure}[tb]
\centering
	\begin{tikzpicture}[scale=0.9, y=-1cm]
		\xdef\x{1}
		\foreach \r/\i in {1/1, 1/2, 1/3, 1/4, 2/1, 2/2, 2/3, 2/4, 3/1, 3/2, 3/3, 3/4}
        {
			\node[anchor=south] at (\x - 0.5, -0.6) {$u_{\r,\i}$};
			\pgfmathparse{\x+1}
			\xdef\x{\pgfmathresult}
		}
        
		\bloomfilter{{$\ge 1$, $\ge 2$, $\ge 3$, $\ge 4$,  $\ge 1$, $\ge 2$, $\ge 3$, $\ge 4$,   $\ge 1$, $\ge 2$, $\ge 3$, $\ge 4$}}{0};

		\belowbrace{0}{4}{0}{Resource 1}
		\belowbrace{4}{8}{0}{Resource 2}
		\belowbrace{8}{12}{0}{Resource 3}        
	\end{tikzpicture}

	\caption{
    	Example of a Bloom filter structure for a problem with 3 resources, each having capacity 4, with $L=U$.
    }
    \label{fig:lossless-bf}
\end{figure}
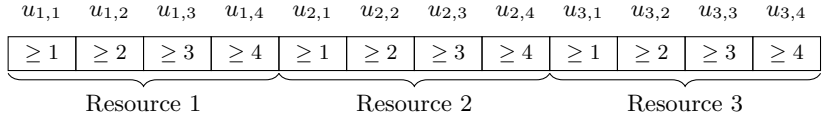 

 Our $L$ building algorithm is as follows:
\begin{enumerate}
	\item
	Start with $L=U$, such as in Figure~\ref{fig:lossless-bf}.
    
    \item \label{step:build-H-test}
 	If $|L|$ is within the prescribed limit, stop.
    
    \item
	Otherwise select $u_{r,k} \in L$ that is least important and delete it.
    Go to step~\ref{step:build-H-test}.
\end{enumerate}
 By `least important' we mean that the deletion of it is expected to have minimal impact of the expected number of false positives.
 Let $L = (\ldots, u_{r,q}, u_{r,k}, u_{r, m}, \ldots)$.
 Consider a job $j$ such that $k \le v_{j,r} < m$ and a slot $t$ such that $q \le A_{t,r} < k$.
 With $L$ as defined above, Bloom filters correctly identify that resources in slot $t$ are insufficient for job $j$: $B^L(j) \not\le B^L(t)$.
 However, without the resource level bit $u_{r,k}$ we get a false positive: $B^L(j) \le B^L(t)$.
 Thus, the probability of false positives caused by deleting $u_{r,k}$ from $L$ in is as follows:
 $$
 \left( \sum_{k = i}^{m - 1} D^r_k \right) \cdot \left( \sum_{k = q}^{i - 1} E^r_k \right) \,,
 $$ 
 where $D^r_k$ is the probability that a randomly chosen job needs exactly $k$ units of resource $r$, and $E^r_k$ is the probability that a certain slot, when we examine it for scheduling a job, has exactly $k$ units of resource $r$ available.
 The probability distribution $D^r$ is produced from the RCPSP instance data during pre-processing.%
 \footnote{In multi-mode extension of RCPSP, this distribution depends on selected modes and hence needs to be obtained empirically, similarly to how we obtain $E^r$.}
 The probability distribution $E^r$ is obtained empirically during the run of \data{} (see Section~\ref{sec:find-enhancements}); each time resource sufficiency is tested within \data{}, its availability is recorded.

\subsection{Additional Speed-ups}
 
 While positive result of a Bloom filter test generally requires further verification using full data, in some circumstances its correctness can be guaranteed.
 In particular, if for some $r \in R$ and $j \in J$ we have $u_{r,k} \in L$ and $v_{j,r} = k$, then the Bloom filter result, whether positive or negative, does not require verification.
 
 Another observation is that updating $B^L(t)$ in $\mathit{update}$ can be done in $O(|R_j|)$ operations instead of $O(|R|)$ operations.
 Indeed, instead of computing $B^L(t)$ from scratch, we can exploit our structure of Bloom filters.
 We update each bit related to resources $r \in R_j$, but we keep intact other bits.
 With some trivial pre-processing, this requires only $O(|R_j|)$ CPU operations.
 
 We also note that if $|R_j| = 1$, i.e.\ job $j$ uses only one resource, then Bloom filters will not speed up the $\mathit{find}$ function for that job and, hence, in such cases we use the standard $\mathit{find}$ function specialised for one resource (see Section~\ref{sec:find-enhancements}).

\section{Hybrid Control Mechanism}
\label{sec:hybrid}
 
 So far we have proposed two improved implementations of SSGS: one using Bloom filters (which we denote \BF{}), and the other one not using Bloom filters (which we denote \NBF{})\@.
 While it may look like \BF{} should always be superior to \NBF{}, in practice \NBF{} is often faster.
 Indeed, Bloom filters usually speed up the $\mathit{find}$ function, but they also slow down $\mathit{update}$, as in \BF{} we need to update not only the values $A_{t,r}$ but also the Bloom filters $B^L(t)$ encoding resource availability.
 If, for example, the RCPSP instance has tight precedence relations and loose resource constraints then $\mathit{find}$ may take only a few iterations, and then the gain in the speed of $\mathit{find}$ may be less than the loss of speed of $\mathit{update}$.
 In such cases \BF{} is likely to be slower then \NBF{}.

 In short, either of the two SSGS implementations can be superior in certain circumstances, and which one is faster mostly depends on the problem instance.
 In this section we discuss how to adaptively select the best SSGS implementation.
 Automated algorithm selection is commonly used in areas such as sorting, where multiple algorithms exist.
 A typical approach is then to extract easy to compute input data features and then apply off-line learning to develop a predictor of which algorithm is likely to perform best, see e.g.~\cite{Guo2003}.
 With sorting, this seems to be the most appropriate approach as the input data may vary significantly between executions of the algorithm.
 Our case is different in that the most crucial input data (the RCPSP instance) does not change between executions of SSGS\@.
 Thus, during the first few executions of SSGS, we can test how each implementation performs, and then select the faster one.
 This is a simple yet effective control mechanism which we call \Hybrid{}.
 
\newcommand{\implementationfont}[0]{\footnotesize}
 
\tikzset{
	cell base/.style={minimum width=1cm, minimum height=1cm, draw=black, transform shape},
	BF/.style={cell base, pattern=north east lines, pattern color=white!60!black, label={[font=\implementationfont]center:BF}},
	NBF/.style={cell base, pattern=north west lines, pattern color=white!80!black, label={[font=\implementationfont]center:NBF}},
	data/.style={cell base, pattern=grid, pattern color=white!80!black, label={[font=\implementationfont]center:data}}
}

\newcommand{\executions}[2]{
	\xdef\x{#2}
	\xdef\y{0}

	\foreach \cell in #1
	{
		\node[\cell] at (\x + 0.5, 0.5+\y) {};
		\pgfmathparse{\x+1}
		\xdef\x{\pgfmathresult}
	}
}

\newcommand{\rightcut}[1]{
	\draw (#1, 0) -- (#1 + 0.3, 0);
	\draw (#1, 1) -- (#1 + 0.3, 1);
	\draw[decorate, decoration=snake] (#1 + 0.3, 0) -- (#1 + 0.3, 1);
}

\newcommand{\leftcut}[1]{
	\draw (#1, 0) -- (#1 - 0.3, 0);
	\draw (#1, 1) -- (#1 - 0.3, 1);
	\draw[decorate, decoration=snake] (#1 - 0.3, 0) -- (#1 - 0.3, 1);
}

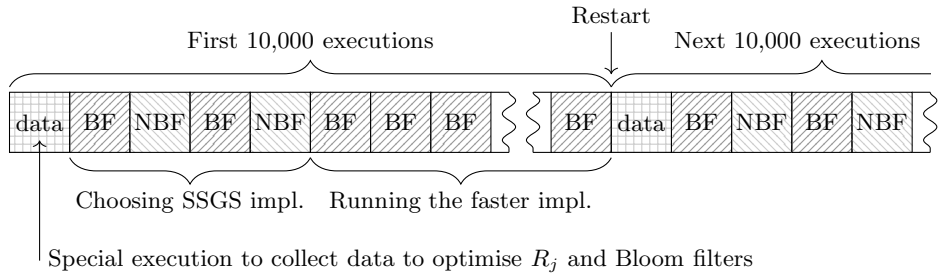
\begin{figure}[tb]
\centering
	\begin{tikzpicture}[scale=0.8, y=-1cm]

		\executions{{data, BF, NBF, BF, NBF, BF, BF, BF}}{0};
        \rightcut{8};
        \leftcut{9};
		\executions{{BF, data, BF, NBF, BF, NBF}}{9};
        \rightcut{15};

		\draw[decorate,decoration={brace,amplitude=10pt,raise=2pt}] (0,0) -- (10,0) node[midway, above=3ex] {First 10,000 executions};
		\draw[decorate,decoration={brace,amplitude=10pt,raise=2pt}] (5,1) -- (1,1) node[midway, below=3ex] {Choosing SSGS impl.};
		\draw[decorate,decoration={brace,amplitude=10pt,raise=2pt}] (10,1) -- (5,1) node[midway, below=3ex] {Running the faster impl.};

		\draw[->] (10, -1) node[above] {Restart} -- (10, -0.3);
		\draw[->] (0.5, 2.8) node[anchor=west] {Special execution to collect data to optimise $R_j$ and Bloom filters} -- (0.5, 0.8);
        
        \path [clip] (0, -1) rectangle (15.3, 1);
		\draw[decorate,decoration={brace,amplitude=10pt,raise=2pt}] (10,0) -- (17,0) node[midway, above=3ex, xshift=-1em] {Next 10,000 executions};
	\end{tikzpicture}

	\caption{
	    Stages of the \Hybrid{} control mechanism.
        Each square shows one execution of SSGS, and the text inside describes which implementation of SSGS is used.
        \data{} is always used in the first execution.
        Further few executions (at most 100) alternate between \BF{} and \NBF{}, with each execution being timed.
        Once sign test shows significant difference between the \BF{} and \NBF{} implementations, the faster one is used for the rest of executions.
        After 10,000 executions, previously collected data is erased and adaptation starts from scratch.
    }
    \label{fig:executions}
\end{figure} 

 \Hybrid{} is entirely transparent for the metaheuristic; the metaheuristic simply calls SSGS whenever it needs to evaluate a candidate solution and/or generate a schedule.
 The \Hybrid{} control mechanism is then intelligently deciding each time which implementation of SSGS to use based on information learnt during previous runs.
 An example of how \Hybrid{} performs is illustrated in Figure~\ref{fig:executions}.
 In the first execution, it uses \data{} to collect data required for both \BF{} and \NBF{}.
 For the next few executions, it alternates between \BF{} and \NBF{}, measuring the time each of them takes.
 During this stage, \Hybrid{} counts how many times \BF{} was faster than the next execution of \NBF{}.
 Then we use the sign test~\cite{cohen1996practical} to compare the implementations.
 If the difference is significant (we use a $5\%$ significance level for the sign test) then we stop alternating the implementations and in future use only the faster one.
 Otherwise we continue alternating the implementations, but for at most 100 executions.
 (Without such a limitation, there is a danger that the alternation will never stop -- if the implementations perform similarly; since there are overheads associated with the alternation and time measurement, it is better to pick one of the implementations and use it in future executions.)
 
 Our decision to use the sign test is based on two considerations: first, it is very fast, and second, it works for distributions which are not normal.
 This makes our approach different from~\cite{Lau2006} where the distributions of runtimes are assumed to be normal.
 (Note that in our experiments we observed that the distribution of running times of an SSGS implementation resembles Poisson distribution.)

 As pointed out in this and previous sections, optimal choices of parameters of the SSGS implementations mostly depend on the RCPSP instance -- which does not change throughout the metaheuristic run; however solution $\pi$ also affects the performance.
 It should be noted though that metaheuristics usually apply only small changes to the solution at each iteration, and hence solution properties tend to change relatively slowly over time.
 Consequently, we assume that parameters chosen in one execution of SSGS are likely to remain efficient for some further executions.
 Thus, \Hybrid{} `restarts' every 10,000 executions, by which we mean that all the internal data collected by SSGS is erased, and learning starts from scratch, see Figure~\ref{fig:executions}.
 This periodicity of restarts is a compromise between accuracy of choices and overheads, and it was shown to be practical in our experiments.

\section{Empirical Evaluation}
\label{sec:evaluation}

 
 In this Section we evaluate the implementations of SSGS discussed above.
 To replicate conditions within a metaheuristic, we designed a simplified version of Simulated Annealing.
 In each iteration of our metaheuristic, current solution is modified by moving a randomly selected job into a new randomly selected position (within the feasible range).
 If the new solution is not worse than the previous one, it is accepted.
 Otherwise the new solution is accepted with 50\% probability.
 Our metaheuristic performs 1,000,000 iterations before terminating.
 
 We evaluate \BF{}, \NBF{} and \Hybrid{}.
 These implementations are compared to `conventional' SSGS, denoted by \Conv{}, which does not employ any preprocessing or Bloom filters and uses conventional implementation of $\mathit{find}$ (Algorithm~\ref{alg:naive}).

 We found that instances in the standard RCPSP benchmark set PSPLIB~\cite{Kolisch1997} occupy a relatively small area of the feature space.
 For example, all the RCPSP instances in PSPLIB have exactly four resources, and the maximum job duration is always set to 10.
 Thus, we chose to use the PSPLIB instance generator, but with a wider range of settings.
 Note that for this study, we modified the PSPLIB instance generator by allowing jobs not to have any precedence relations.
 This was necessary to extend the range of network complexity parameter (to include instances with scarce precedence relations), and to speed up the generator, as the original implementation would not allow us to generate large instances within reasonable time.

 All the experiments are conducted on a Windows Server 2012 machine based on Intel Xeon E5-2690 v4 2.60~GHz CPU.
 No concurrency is employed in any of the implementations or tests.
 
 To see the effect of various instance features on SSGS performance, we select one feature at a time and plot average SSGS performance against the values of that feature.
 The rest of the features (or generator parameters) are then set as follows: number of jobs 120, number of resources 4, maximum duration of job 10, network complexity 1, resource factor 0.75, and resource strength 0.1.
 These values correspond to some typical settings used in PSPLIB\@.
 For formal definitions of the parameters we refer to \cite{Kolisch1997}.
 
 For each combination of the instance generator settings, we produce 50 instances using different random generator seed values, and in each of our experiments the metaheuristic solves each instance once.
 Then the runtime of SSGS is said to be the overall time spent on solving those 50 instances, over 50,000,000 (which is the number of SSGS executions).
 The metaheuristic overheads are relatively small and are ignored.

\pgfplotsset{Hybrid line/.append style={line width=5pt, black, opacity=0.3}}
\pgfplotsset{BF line/.append style={black, mark=x, mark size=3pt}}
\pgfplotsset{Non-BF line/.append style={thick, black, mark=*}}

\newcommand{\varyparam}[1]{
	\addplot[BF line] table[
		x=RS,
		y expr=100 * \thisrow{BF} / \thisrow{Conventional},
	] {#1};
    \label{plots:BF}

	\addplot[Non-BF line] table[
		x=RS,
		y expr=100 * \thisrow{Non-BF} / \thisrow{Conventional},
	] {#1};
    \label{plots:NBF}

	\addplot[Hybrid line] table[
		x=RS,
		y expr=100 * \thisrow{Hybrid} / \thisrow{Conventional},
	] {#1};
    \label{plots:Hybrid}
}

\begin{figure}[h!]
\begin{tikzpicture}
    \begin{groupplot}[
    		group style={
            	group name=myplot,
	            group size=2 by 3,
                vertical sep=1.6cm},
            height=5.5cm,
            width=6.6cm,
            grid=major,
            ylabel style={yshift=-1em}]
        \nextgroupplot[xlabel={Number of jobs}, ylabel={Runtime, \%}]
			\varyparam{jobs-number.dat}
        \nextgroupplot[xlabel={Number of resources}]
			\varyparam{resources-number.dat}
        \nextgroupplot[xlabel={Resource strength}, ylabel={Runtime, \%}]
			\varyparam{resource-strength.dat}
        \nextgroupplot[xlabel={Resource factor}]
			\varyparam{resource-factor.dat}
        \nextgroupplot[xlabel={Network complexity}, ylabel={Runtime, \%}]
			\varyparam{network-complexity.dat}
        \nextgroupplot[xlabel={Maximum job duration}]
			\varyparam{duration.dat}
    \end{groupplot}

\path (myplot c1r1.north west|-current bounding box.north)--
      coordinate(legendpos)
      (myplot c2r1.north east|-current bounding box.north);
\matrix[
    matrix of nodes,
    anchor=south,
    draw,
    inner sep=0.2em,
    draw
  ]at([yshift=2ex]legendpos)
  {
    \ref{plots:BF} & \BF{} & [5pt]
    \ref{plots:NBF} & \NBF{} & [5pt]
    \ref{plots:Hybrid} & \Hybrid{} \\
  };
\end{tikzpicture}

\caption{
	These plots show how performance of the SSGS implementations depends on various instance features.
    Vertical axis gives the runtime of each implementation relative to \Conv{}.
    (\Conv{} graph would be a horizontal line $y = 100\%$.)}
\label{fig:performance}
\end{figure}
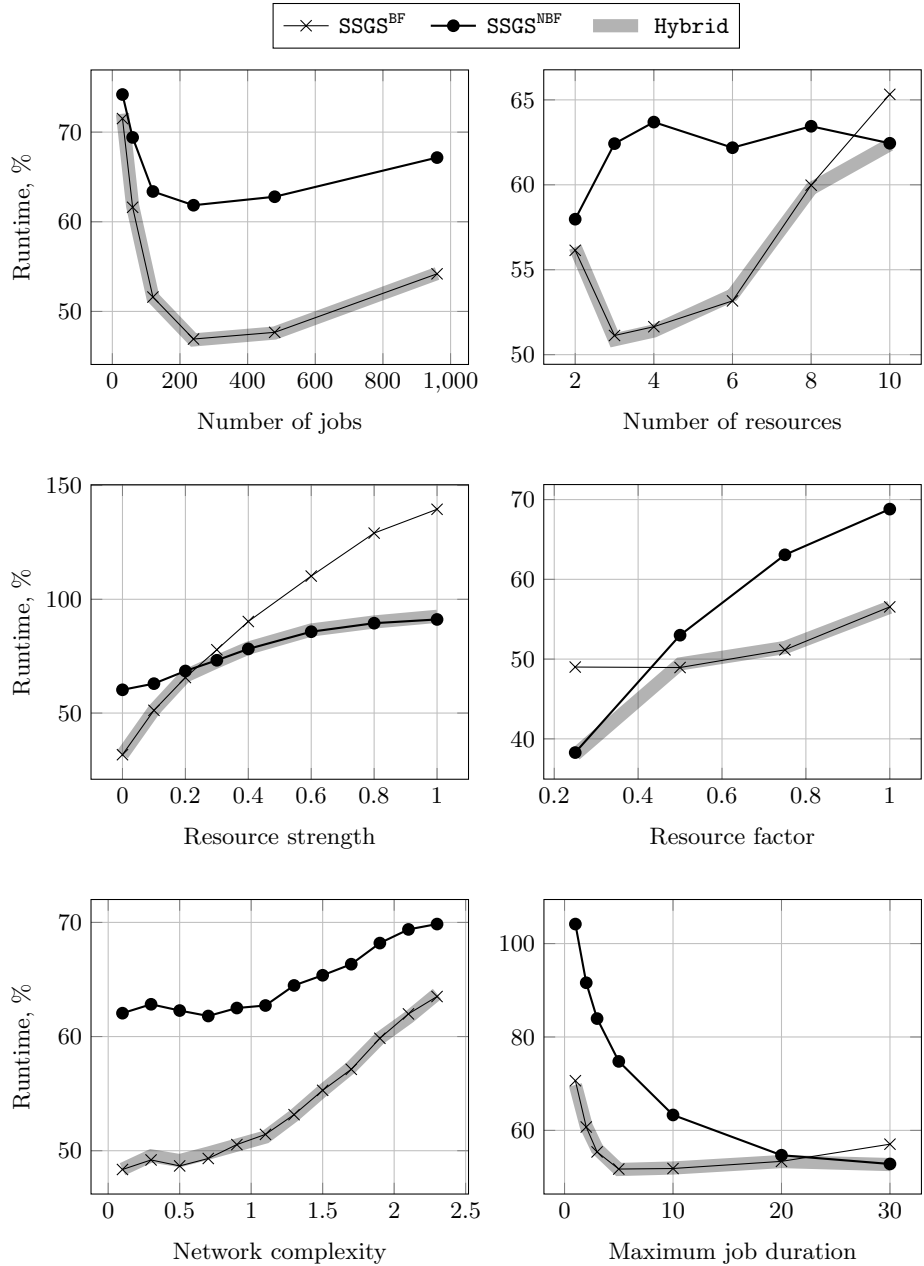

 From the results reported in Figure~\ref{fig:performance} one can see that our implementations of SSGS are generally significantly faster than \Conv{}, but performance of each implementation varies with the instance features.
 In some regions of the instance space \BF{} outperforms \NBF{}, whereas in other regions \NBF{} outperforms \BF{}.
 The difference in running times is significant, up to a factor of two in our experiments.
 At the same time, \Hybrid{} is always close to the best of \BF{} and \NBF{}, which shows efficiency of our algorithm selection approach.
 In fact, when \BF{} and \NBF{} perform similarly, \Hybrid{} sometimes outperforms both; this behaviour is discussed below.

 Another observation is that \NBF{} is always faster than \Conv{} (always below the 100\% mark) which is not surprising; indeed, \NBF{} improves the performance of both $\mathit{find}$ and $\mathit{update}$.
 In contrast, \BF{} is sometimes slower than \Conv{}; on some instances, the speed-up of the $\mathit{find}$ function is overweighed by overheads in both $\mathit{find}$ and $\mathit{update}$.
 Most important though is that \Hybrid{} outperforms \Conv{} in each of our experiments by 8 to 68\%, averaging at 43\%.
 In other words, within a fixed time budget, an RCPSP metaheuristic employing \Hybrid{} will be able to run around 1.8 times more iterations than if it used \Conv{}.







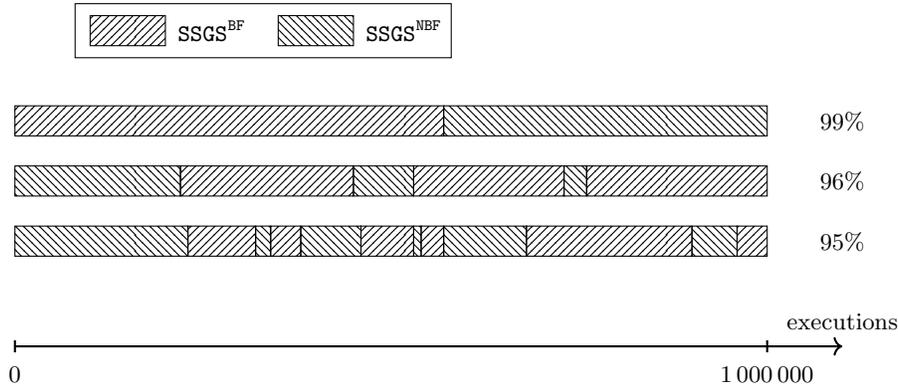
\begin{figure}[tb]
\tikzset{
	BF/.style={pattern=north east lines},
	NBF/.style={pattern=north west lines},
}
\begin{tikzpicture}[
    xscale=0.1,
    yscale=0.4
]

\draw[NBF] (0, 1) rectangle (23, 2);
\draw[BF] (23, 1) rectangle (32, 2);
\draw[NBF] (32, 1) rectangle (34, 2);
\draw[BF] (34, 1) rectangle (38, 2);
\draw[NBF] (38, 1) rectangle (46, 2);
\draw[BF] (46, 1) rectangle (53, 2);
\draw[NBF] (53, 1) rectangle (54, 2);
\draw[BF] (54, 1) rectangle (57, 2);
\draw[NBF] (57, 1) rectangle (68, 2);
\draw[BF] (68, 1) rectangle (90, 2);
\draw[NBF] (90, 1) rectangle (96, 2);
\draw[BF] (96, 1) rectangle (100, 2);
\node at (110, 1.5) {95\%};

\draw[NBF] (0, 3) rectangle (22, 4);
\draw[BF] (22, 3) rectangle (45, 4);
\draw[NBF] (45, 3) rectangle (53, 4);
\draw[BF] (53, 3) rectangle (73, 4);
\draw[NBF] (73, 3) rectangle (76, 4);
\draw[BF] (76, 3) rectangle (100, 4);
\node at (110, 3.5) {96\%};

\draw[BF] (0, 5) rectangle (57, 6);
\draw[NBF] (57, 5) rectangle (100, 6);
\node at (110, 5.5) {99\%};


\draw[thick, ->] (0, -2) -- (110, -2) node[anchor=south, above=3pt] {executions};
\draw[thick] (0, -1.8) -- (0, -2.2) node[anchor=north, below=2pt] {0};
\draw[thick] (100, -1.8) -- (100, -2.2) node[anchor=north, below=2pt] {1\,000\,000};

\draw[BF] (10, 8) rectangle (20, 9);
\node[anchor=west] at (20.5, 8.5) {\BF};
\draw[NBF] (35, 8) rectangle (45, 9);
\node[anchor=west] at (45.5, 8.5) {\NBF};
\draw (8, 7.6) rectangle (58, 9.4);

\end{tikzpicture}
\caption[Diagram]{
	This diagram shows how \Hybrid{} switches between SSGS implementations while solving three problem instances.  The number on the right shows the time spent by \Hybrid{} compared with the time that would be needed if only the implementation chosen at the start would be used for all iterations.
}
\label{fig:adaptive}
\end{figure}

 To verify that \Hybrid{} exhibits the adaptive behaviour and does not just stick to whichever implementation has been chosen initially, we recorded the implementation it used in every execution for several problems, see Figure~\ref{fig:adaptive}. 
 For this experiment, we produced three instances: first instance has standard parameters except Resource Strength is 0.2; second instance has standard parameters except Resource Factor is 0.45; third instance has standard parameters except Maximum Job Duration is 20.
 These parameter values were selected such that the two SSGS implementations would be competitive and, therefore, switching between them would be a reasonable strategy. 
 One can see that the switches occur several times throughout the run of the metaheuristic, indicating that \Hybrid{} adapts to the changes of solution.
 For comparison, we disabled the adaptiveness and measured the performance if only implementation chosen initially is used throughout all iterations; the results are shown on Figure~\ref{fig:adaptive}.
 We conclude that \Hybrid{} benefits from its adaptiveness.

\section{Conclusions and Future Work}

 In this paper we discussed the crucial component of many scheduling metaheuristics, the serial schedule generation scheme (SSGS).
 SSGS eats up most of the runtime in a typical scheduling metaheuristic, therefore performance of SSGS is critical to the performance of the entire metaheuristic, and thus each speed-up of SSGS has significant impact.
 We described existing and some new speed-ups to SSGS, including preprocessing and automated parameter control.
 This implementation clearly outperformed the `conventional' SSGS in our experiments.
 We further proposed a new implementation that uses Bloom filters, particularly efficient in certain regions of the instance space.
 To exploit strengths of both implementations, we proposed a hybrid control mechanism that learns the performance of each implementation and then adaptively chooses the SSGS version that is best for a particular problem instance and phase of the search.
 Experiments showed that this online algorithm selection mechanism is effective and makes \Hybrid{} our clear choice.
 Note that \Hybrid{} is entirely transparent for the metaheuristic which uses it as if it would be simple SSGS; all the learning and adaptation is hidden inside the implementation.

 The idea behind online algorithm selection used in this project can be further developed by making the number of executions between restarts adaptable.
 To determine the point when the established relation between the SSGS implementations may have got outdated, we could treat the dynamics of the implementations' performance change as two random walks, and use the properties of these two random walks to predict when they may intersect.
 
 All the implementations discussed in the paper, in C++, are available for downloading from \url{http://csee.essex.ac.uk/staff/dkarap/rcpsp-ssgs.zip}.
 The implementations are transparent and straightforward to use.
 
 While we have only discussed SSGS for the simple RCPSP, our ideas can easily be applied in RCPSP extensions.
 We expect some of these ideas to work particularly well in multi-project RCPSP, where the overall number of resources is typically large but only a few of them are used by each job. 

\bigskip

\noindent
\textbf{Acknowledgements} 
 We would like to thank Prof.~Rainer Kolisch for providing us with a C++ implementation of the PSPLIB generator.
 It should be noted that, although the C++ implementation was developed to reproduce the original Pascal implementation, the exact equivalence cannot be guaranteed; also, in our experiments we used a modification of the provided C++ code.

\bibliographystyle{plain}
\bibliography{refs}

\end{document}